\pdfoutput=1

\documentclass[twocolumn]{article}

\usepackage{amsmath}
\usepackage{graphicx}
\usepackage{sistyle}
\usepackage{bm}
\usepackage{dsfont}
\usepackage{epsfig}
\usepackage{feynmf}
\usepackage{blindtext, rotating}
\usepackage{mathtools}
\usepackage{dsfont}
\usepackage{subcaption}
\usepackage{physics}
\usepackage{amsfonts}
\usepackage{ragged2e}
\usepackage{authblk}
\usepackage{abstract}
\newcommand{\overbar}[1]{\mkern 1.5mu\overline{\mkern-1.5mu#1\mkern-1.5mu}\mkern 1.5mu}
\usepackage{xcolor} 

\DeclareCaptionJustification{justified}{\justifying}

\newcommand{\me}{\operatorname{e}}
\usepackage[margin=0.75in]{geometry}

\begin{document}

\title{Entanglement Dynamics in Dispersive Optomechanics: \linebreak Non-Classicality and Revival}
\author{Igor Brand\~ao\thanks{igorbrandao@aluno.puc-rio.br}}
\author{Bruno Suassuna\thanks{bruno.b.suassuna@gmail.com}}
\author{Bruno Melo\thanks{brunomelo@aluno.puc-rio.br}}
\author{Thiago Guerreiro\thanks{barbosa@puc-rio.br}}

\affil{Department of Physics, Pontifical Catholic University of Rio de Janeiro, Rio de Janeiro 22451-900, Brazil}
\date{}

\twocolumn[
    \maketitle
    \begin{onecolabstract}
We study entanglement dynamics in dispersive optomechanical systems consisting of two optical modes and a mechanical oscillator inside an optical cavity. 
The two optical modes interact with the mechanical oscillator, but not directly with each other. The appearance of optical
entanglement witnesses non-classicality of the oscillator. We study the dependence of the entanglement dynamics with the optomechanical coupling, the mean photon number in the cavity and the oscillator temperature. 
An experimental
realization with ultracold atomic ensembles is proposed.
 \end{onecolabstract}
]
\saythanks



\section{Introduction} 
Entanglement is one of the most striking phenomena of quantum theory \cite{Schroedinger1935}. Generating, manipulating and measuring entanglement in systems with many constituents and with a large number of degrees of freedom is one of the challenges of Quantum Information and Metrology \cite{Vedral2008}, and an interesting frontier in Fundamental Physics \cite{Preskill2018}. In particular, entangling massive objects could open the way to interesting tests of quantum theory \cite{Feynman1996, Diosi1987} and experiments aimed at probing gravitational effects of quantum mechanical matter \cite{Penrose1996, Bose2017, Blencowe2013, Pikovski2015, Belenchia2019, Carney2019}. Optomechanical systems provide a resourceful platform to this end. 


It is well known that entanglement of massive objects can be realized in quantum cavity optomechanical experiments \cite{Asplemeyer2014}. 
For instance, a cavity with a moving end mirror can be used to generate entangled ``cat states'' of both light \cite{Mancini1997, Giovannetti2001} and matter \cite{Bose1997, Mancini2002, Zhang2003, Pinrandola2006}. 
Similar systems have also been proposed as an effective nonlinear medium \cite{Corbitt2006, Wipf2008, XinYou2015} and squeezing \cite{Brooks2012, Purdy2013, Aggarwal2018} as well as optical entanglement \cite{Chen2020} have been experimentally demonstrated in a variety of set-ups such as cavity cold atomic ensembles \cite{Murch2008}, dispersive dielectric membranes \cite{Jayich2008} and silicon micro-resonators \cite{Painter2013}. In the linearized regime, particularly, entanglement dynamics \cite{Tian2013, Rakhubovsky2020} and stationary entanglement \cite{Barzanjeh2019, Wang2013, Gut2020, Ockeloen-Korppi2018} have attracted much attention as these systems have a wide applicability ranging from precision force measurements \cite{Palomaki2013, Ockeloen-Korppi2018} to quantum networks \cite{Riedinger2016, Riedinger2018}.

Certifying quantumness of optomechanical systems, however, is a far-from-obvious task. Relations among entanglement and non-classicality measures of quantum states can be used to probe the quantum nature of innaccessible objects such as a harmonic oscillator in an optical cavity \cite{Krisnanda2017}. Recurrence of optical squeezing in a cavity with a moving mirror has also been proposed as a witness of non-classicality \cite{Ma2020} and it has been shown that when two subsystems locally interact with a third one, but not directly with each other, the appearance of entanglement among those subsystems is sufficient to prove non-classicality of the third party \cite{Marletto2017}.
Building on some of these ideas the present work studies the entanglement dynamics of a dispersive optomechanical system, how to use that dynamics to probe the quantum nature of the oscillator through optical degrees of freedom and how to optimize the generated optical entanglement by careful choice of the optomechanical coupling and the number of photons in an experiment.

\begin{figure}[t!]
    \centering
    \includegraphics[scale = 0.7]{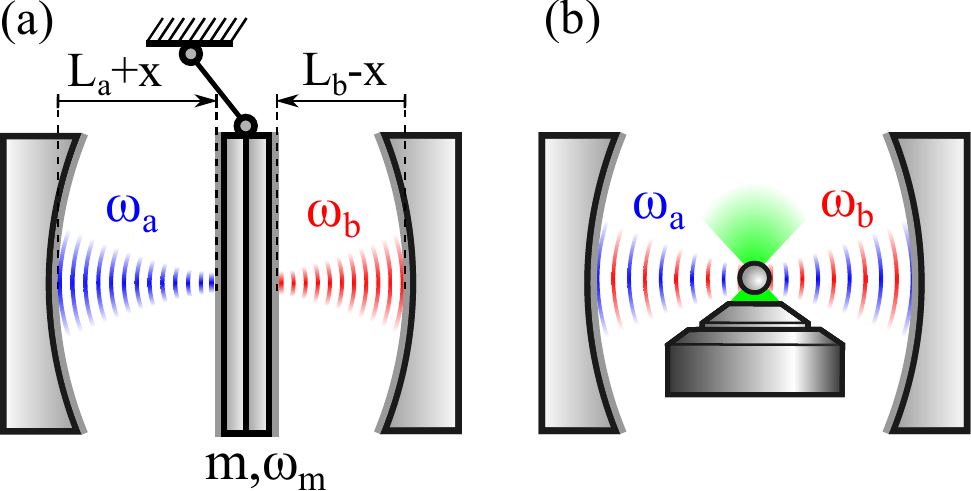}
    \caption[]{\small(a) Schematics of coupled optical cavities sharing a ``mirror-in-the-middle'' under a harmonic potential. No photon transfer from one cavity to the other is allowed. 
    (b) Schematics of a particle trapped by an optical tweezer coupled to  two  modes  of  a  cavity. The particle can be considered as a Silica nano-sphere or a cloud of ultracold atoms. When the levitated object is properly positioned, the Hamiltonian describing both systems acquires the same form. 
    }
    \label{fig:Figure1}
\end{figure}
Considering as possible implementations levitated optomechanical systems, such as Silica nano-spheres or cold atomic ensembles, and a ``two-sided'' cavity with a moving mirror in the middle, we map how entanglement appears and evolves among the various optical and mechanical subsystems for different optomechanical coupling strengths and optical field intensities. 
The appearance of mechanically induced optical entanglement and its subsequent death and revival are generic in these systems, and thus could be used to probe the quantization of the center-of-mass of the moving object in real experiments.  
We also point out in a simplified context that under certain circumstances entanglement seems to ``flow'' through different subsystems, and such dynamics can be used to infer non-classicality and entanglement among different components of the system. 
We consider examples of both non-Gaussian and Gaussian initial quantum states, for which we study the dynamics of concurrence and the Duan criteria \cite{DuanCriteria}, respectively. 
An experiment using levitated cold atomic ensembles is proposed.

\section{Hamiltonian description}\label{sec:Hamiltonian}
The system we are primarily interested in is shown in Figure \ref{fig:Figure1}(a): two optical cavities, of lengths $L_a$ and $L_b$, are populated by modes of frequencies $\omega_a$ and $\omega_b$ and share a common perfect movable mirror of mass $m$ subject to a harmonic potential of frequency $\omega_m$. We refer to this as the ``mirror-in-the-middle'' configuration. 
In this system the optical modes never interact directly, except via the dispersive coupling due to the presence of the mechanical mode. Since we are interested in studying entanglement dynamics in optomechanics, we shall assume the cavities can be initialized in particular states and the laser driving-term can be turned off during the course of the experiment.
It is also assumed that optical losses are negligible during the time of the experiment and a discussion of the conditions under which this is true and the experimental feasibility is addressed in the experimental proposal.

The Hamiltonian of the system reads \cite{Asplemeyer2014}
\begin{align}
\frac{H}{\hbar}=&\,
\omega_a\hat{a}^\dagger \hat{a}+\omega_b\hat{b}^\dagger \hat{b}+\omega_m\hat{c}^\dagger \hat{c}\nonumber\\ 
&-g_{0,a} \hat{a}^\dagger \hat{a}(\hat{c}^\dagger+\hat{c}) +g_{0,b} \hat{b}^\dagger \hat{b}(\hat{c}^\dagger+\hat{c})\, ,
    \label{eq:Hamiltonian}
\end{align}
where $g_{0,i}=\omega_ix_{zpf}/L_i$ are the optomechanical couplings, with $x_{zpf}=\sqrt{\hbar/2m\omega_m}$ the zero point fluctuation of the mirror and $\hat{a}$, $\hat{b}$, $\hat{c}$ ($\hat{a}^\dagger$, $\hat{b}^\dagger$, $\hat{c}^\dagger$) are the annihilation (creation) operators of each optical and mechanical modes, denoted by $ A $, $B $ and $ C $, respectively. Such Hamiltonian can also be implemented using a cavity with a levitated nano-particle \cite{Chang2009, Romero_Isart_2010, Kiesel2013} or an ultracold atom cloud \cite{Murch2008, Neumeier2018, Brennecke2008} properly positioned along the cavity axis. This is illustrated in Figure \ref{fig:Figure1}(b) (see Appendices \ref{appendix:levitated_implementation} and \ref{appendix:ultracol_atom_implementation} for details).

Assuming equal frequencies for the optical modes $\omega_a = \omega_b$, and approximately equal cavity lengths $L_a \sim L_b$ we may simplify the notation and directly write $g_0 \equiv g_{0,a} \sim  g_{0,b}$. The unitary evolution operator resulting from equation \eqref{eq:Hamiltonian} becomes
\begin{align}
\label{eq:Evolution Operator}
    \hat{U}(t) = &\me^{-i r_a \hat{a}^\dagger \hat{a} t}  \me^{-i r_b \hat{b}^\dagger \hat{b} t} \me^{-i B(t) (\hat{a}^\dagger\hat{a} - \hat{b}^\dagger\hat{b})^2}\nonumber\\& \me^{+ k\hat{a}^\dagger\hat{a}(\eta \hat{c}^\dagger-\eta^*\hat{c})}\me^{- k \hat{b}^\dagger\hat{b}(\eta \hat{c}^\dagger-\eta^*\hat{c})} \me^{-i\hat{c}^\dagger \hat{c} t}\,,
\end{align}
where we define the dimensionless optomechanical coupling $k=g_0/\omega_m$,  the ``normalized frequencies'' $r_i=\omega_i/\omega_m$, the scaled time $\omega_m t \rightarrow t$, and the functions $\eta(t) = 1-e^{-it}$ and $B(t) = -k^{2} (t - \sin{t}) $. Note the evolution operator is comprised of a \textit{Kerr-like} term, responsible for an effective optical non-linearity \cite{Takatsuji1967}, as well as an \textit{optically-driven} displacement operator acting on the mechanical mode. 

It is expected that a generic separable state will evolve into an entangled one by virtue of the unitary evolution \eqref{eq:Evolution Operator}. We note that if an initially separable state gives birth to optical entanglement then there will certainly be its entanglement death. This springs from the fact that when $B(t) = 2\pi n, n\in\mathbb{N}$, the term in the evolution operator responsible for entangling the optical modes reduces to the identity operator at those times, therefore preserving the separability of the initial state. Analogous arguments show that opto-mechanical entanglement must also face death when $\eta(t)=0$. 


Not every state will evolve to an entangled one, as can be seen by considering the energy eigenstates of the system
\begin{equation}
    \hat{\mathcal{D}}_C(k(n_A-m_B))\ket{n_A,m_B,\ell_C} ,
\end{equation}
where $\lbrace\ket{n_A,m_B,\ell_C}\rbrace$ denotes the number basis and $\hat{\mathcal{D}}_C(\alpha)$ the displacement operator acting on the mechanical oscillator, mode $C$, by a displacement $\alpha\in\mathbb{C}$ (see Appendix \ref{appendix:unitary_evolution} for details).







	




\section{Qubit states}
Consider the cavities in Figure \ref{fig:Figure1}(a) initially populated by the state
\begin{eqnarray}
\vert \Psi (0) \rangle = \left( \dfrac{\vert 0 \rangle + \vert 1 \rangle}{\sqrt{2}} \right) \otimes \left(  \dfrac{\vert 0 \rangle + \vert 1 \rangle}{\sqrt{2}} \right) \otimes \vert 0 \rangle.
\label{eq:non-gauss}
\end{eqnarray}
We refer to these as `qubit states' as they are restricted to the vacuum-one-photon subspace. These states can be prepared in quantum optics in an approximate way using photon pair sources and displacement-based detection \cite{Lombardi2002, Guerreiro2016} or non-linear light-matter interactions in cavity quantum electrodynamics \cite{Peyronel2012, Hofheinz2009}. 
In levitated quantum electrodynamics \cite{Hornberger2020}, it is also possible to couple two qubits to a nano-sphere or rotor according to the interaction Hamiltonian \eqref{eq:Hamiltonian}, with the qubits assuming the role of the optical fields and the levitated object the role of the mirror-in-the-middle. Among the advantages of this type of scheme is the fact that the optomechanical coupling admits a wide tunability, potentially allowing tests of the optomechanical interaction in novel regimes. Moreover, read-out of the ``optical'' modes can be achieved through standard qubit read-out techniques \cite{Krantz2019}.
The ``mirror-in-the-middle'' is taken to be in the ground state for simplicity; in the next section we shall consider the effects of a finite temperature oscillator. 
Notice the initial state is separable and hence the appearance of entanglement between modes $ A $ and $ B $ would evidence the non-classical nature of mode $ C $ \cite{Marletto2017}. 
Time evolution of \eqref{eq:non-gauss} in the interaction picture is explicitly given by 
\begin{align}
    \vert \Psi(t) \rangle  = \dfrac{ \vert 00 \rangle }{2}\vert 0 \rangle + e^{iB(t)} \dfrac{ \vert 01 \rangle }{2} \hat{\mathcal{D}}_C(k\xi(t)) \vert 0 \rangle + \nonumber \\
    + e^{iB(t)} \dfrac{ \vert 10 \rangle }{2}  \hat{\mathcal{D}}_C(-k\xi(t)) \vert 0 \rangle + \dfrac{ \vert 11 \rangle }{2}  \vert 0 \rangle  , \label{eq:evolved_state}
\end{align}
where 
$\xi(t)= e^{it}\eta(t)$. The evolved state \eqref{eq:evolved_state} exhibits entanglement between modes $A$ and $B$. This can be promptly seen by noticing that coherent states are non-orthogonal and, in the limit of small coupling $ k $, the state assumes the form $\vert \Psi(t) \rangle \simeq \vert \varphi_{AB} \rangle \otimes \vert \varphi_{C} \rangle $, where
\begin{eqnarray}
\vert \varphi_{AB} \rangle & \simeq & \dfrac{\vert 0 \rangle}{2}    \left( \vert 0 \rangle  + e^{iB(t)} \vert 1 \rangle \right)  \nonumber \\
   &+& e^{iB(t)} \dfrac{\vert 1 \rangle}{2} \left( \vert 0 \rangle  + e^{-iB(t)}\vert 1 \rangle   \rangle  \right )    
\end{eqnarray}
and $ \vert \varphi_{C} \rangle \simeq  \vert 0 \rangle $. For times $t$ such that $\vert B(t)\vert \simeq \pi/2 + 2\pi n, n\in \mathbb{N}$, the state $ \vert \varphi_{AB} \rangle $ becomes maximally entangled. 
The origin of this entanglement can be heuristically explained by a simple argument: the ground state of the mechanical oscillator is a Gaussian wave packet in the position basis. Each possible position adds-up coherently introducing correlations in the lengths of the left and right cavities in Figure \ref{fig:Figure1}(a). This imprints correlations in the phases of the corresponding electromagnetic fields in modes $ A $ and $ B $ giving rise to entanglement. As long as the dimensionless optomechanical coupling $ k $ due to radiation pressure on the middle mirror is sufficiently small, mode $ C $ will be approximately unperturbed and therefore, to a good approximation, disentangled from $ AB $. On the other hand, as the coupling strength increases mode $ C $ can become significantly entangled with modes $ A $ and $ B$; the question of entanglement among different subsystems as a function of $ k $ will be addressed in the next section.

To quantitatively evaluate the entanglement in \eqref{eq:evolved_state} we calculate the three-partite density matrix $ \rho_{ABC} = \vert \Psi(t) \rangle \langle \Psi(t) \vert  $,  from which we obtain the reduced state $ \rho_{AB} = \mathrm{Tr}_{C} \left( \rho_{ABC} \right)  $:
\begin{align}
\hspace*{-0.2em}\rho_{AB}(t) = \dfrac{1}{4} 
\begin{pmatrix}
1 & \me^{C(t)} & \me^{C(t)} & 1 \\
\me^{C^*\hspace*{-0.15em}(t)} & 1 &
\me^{- 2k^{2} \vert \eta(t) \vert^{2}}  & \me^{C^*\hspace*{-0.15em}(t)}  \\
\me^{C^*\hspace*{-0.15em}(t)} & \me^{- 2k^{2} \vert \eta(t) \vert^{2}} &
1  & \me^{C^*\hspace*{-0.15em}(t)}  \\
1 & \me^{C(t)} & \me^{C(t)} & 1 
\end{pmatrix}
\label{eq:toy_matrix}
\end{align}
with $C(t)=iB(t) -k^{2} \vert \eta(t) \vert^{2}/2$. Notice that for small values of $ k $ the mirror is ``weakly entangled'' with modes $ A $ and $ B $ and some of the off-diagonal terms of the reduced density matrix acquire exponentials that alternate between periods of decay and periods of growth. This can be seen as an example of a weak form of decoherence and ``non-Markovian'' evolution for the partitions of the whole system, in which information about the optical modes leak into correlations with the mirror and is later retrieved. The mirror introduces a ``memory'' in the system \cite{HeinzPeterBreuer2007}. Non-Markovianity springs from the fact that the mirror is part of the system under study and hence its degrees of freedom are under control. 

\begin{figure}[t]
    \centering
    \includegraphics[width=8.6cm]{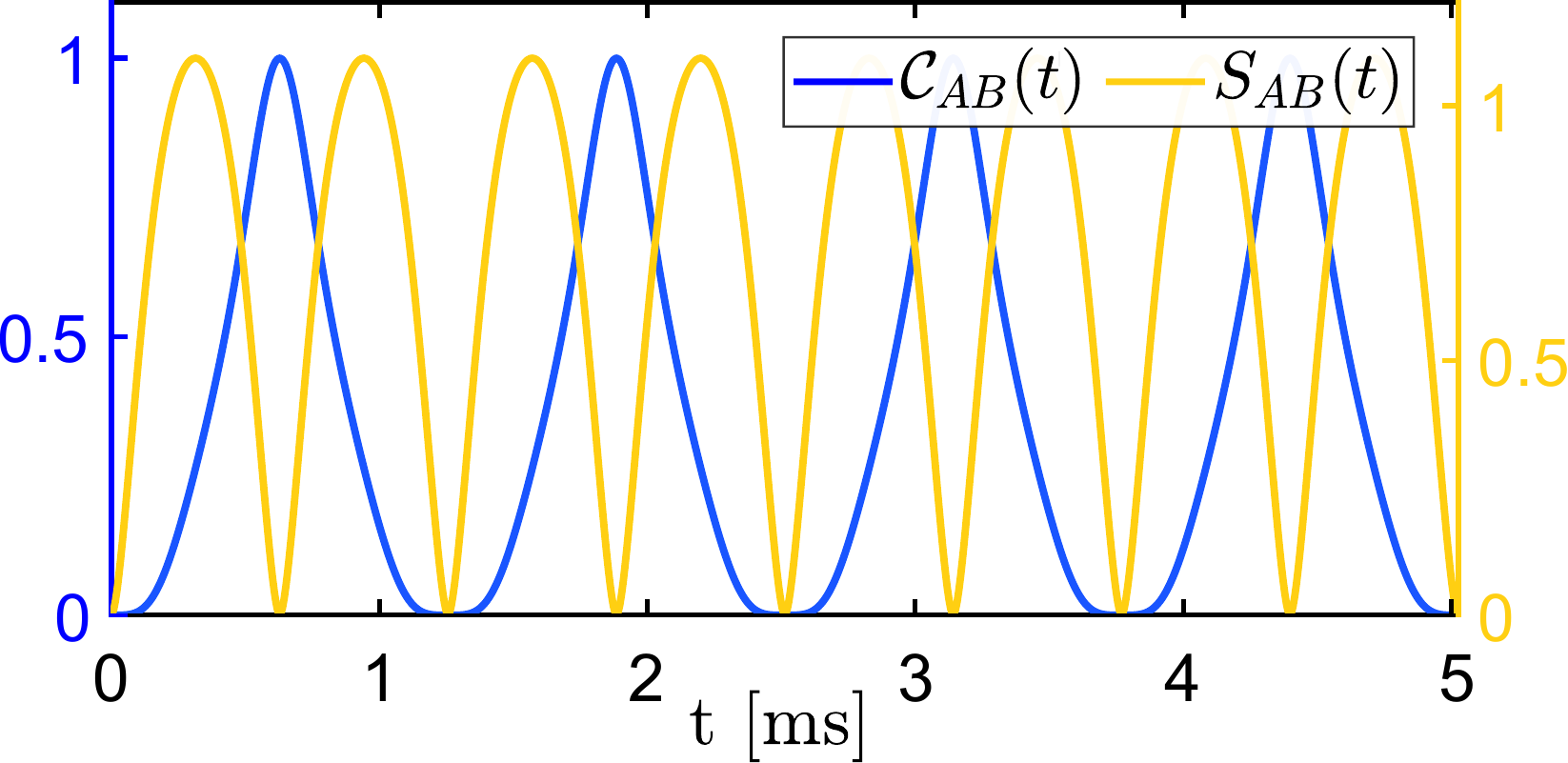}
    \caption[]{\small Concurrence (blue) and von Neumann entropy (yellow) for $ \rho_{AB}(t) $ as a function of time. For this plot  $k = 0.5$.} 
    \label{fig:Fgiure2}
    \label{}
\end{figure}

Since the state $ \rho_{AB}(t) $ is restricted to the subspace spanned by $\lbrace\vert 0 \rangle, \vert 1 \rangle\rbrace$, we can use concurrence as a measure of entanglement. 
As an example, plots of the concurrence $ \mathcal{C}_{AB}(t) $ and von Neumann entropy $ S_{AB}(t) $ of $ \rho_{AB} $ are shown in Figure~\ref{fig:Fgiure2} for coupling value $ k = 0.5 $. 
The optical modes exhibit positive concurrence and hence entanglement as a function of time. Moreover, the system exhibits sudden death and birth of entanglement. 
It is also possible to see that the entropy, which is initially zero, oscillates as a function of time. This is another indication of the non-Markovian nature of the system.
A non-zero entropy of $ AB $ signals entanglement among the three-partite system $ ABC$. 
Moreover, note that the maxima of concurrence (entanglement of $ AB $) coincide with the minima of entropy (entanglement of $ ABC$). This suggests that, after emerging in the system, entanglement ``flows'' (during the limited period of its lifetime) among different partitions of the system. 

\section{Continuous Variable states and finite temperature}
We now consider a scenario in which initially the optical modes are populated by monochromatic coherent states and the moving object (sphere, cloud of atoms or mirror) is in a thermal state at temperature $T$
\begin{align}
    \rho(0)=\ket{\alpha}\bra{\alpha} \otimes \ket{\beta}\bra{\beta} \otimes \frac{1}{Z} \sum_n e^{-\frac{n\hbar\omega_c}{k_B T}} \vert n \rangle \langle n \vert,\ \label{initial_state}
\end{align}
where $Z = \sum_n e^{-\frac{n\hbar\omega_c}{k_B T}} $ is the thermal partition function. Note that although the initial state here considered is Gaussian, the Hamiltonian \eqref{eq:Hamiltonian} has cubic terms in creation and annihilation operators and, therefore, does not preserve Gaussianity \cite{Gaussian_Quantum_Information}. In order to study the dynamics of entanglement for Continuous Variable states, we turn our attention to the time-dependent Duan Criteria \cite{DuanCriteria}. 

We define $D_{ij}$ as half of the Einstein-Podolski-Rosen (EPR) variance for modes $i,j$
\begin{align}
    D_{ij} \equiv \frac{1}{2} \bigg[ \big(\Delta\hat{u}_{ij}\big)^2 + \big(\Delta\hat{v}_{ij}\big)^2 \bigg] \label{eq:EPR_variance}
\end{align}
\noindent where
\begin{align}
    \hat{u}_{ij} &= \hat{x}_{i} + \hat{x}_j \, , \\
    \hat{v}_{ij} &= \hat{p}_{i} - \hat{p}_j 
\end{align}
\noindent are the EPR operators for different modes, $i, j=A,B,C$. The Duan Criteria states that any separable state satisfies 
\begin{equation}
    D_{ij} \geq 1\, .
\end{equation}
\noindent Therefore, if at any time $t$ a violation of the above inequality is observed, modes $i$ and $j$ are necessarily entangled at that time. As the Duan criteria is written in terms of field quadratures, it can be promptly measured with homodyne detection techniques readily available in the laboratory \cite{Davidovich}. 
We also note that when $t=2\pi n/ \omega_m, n\in\mathbb{N}$, the unitary evolution \eqref{eq:Evolution Operator} acts as a two-mode squeezing operator analogous to a $\chi^{(3)}$ interaction in nonlinear optics \cite{Lloyd1999}. In these moments, the optomechanical system behaves as a nonlinear optical source of squeezing, from which quantum correlations can be readout from the leaking fields of the cavity \cite{Kippenberg2004}. 

Given the time evolution operator in equation \eqref{eq:Evolution Operator} we are able to find analytical expressions for the EPR variance of every bipartition of the system (see Appendix \ref{appendix:Duan_Criteria} for details). For the optical modes we have

\begin{equation}
   \begin{split}
    D_{AB}(t) = 1 + \bigg[|\alpha|^2 +  2\alpha\beta \cos\big((r_a+r_b)t\big) + |\beta|^2 \bigg] \\ - \bigg[|\alpha|^2 + 2\alpha\beta\cos\big( (r_a+r_b)t + 2B(t) \big) + |\beta|^2\bigg]\\ \times  \me^{-2\big[\vert\alpha\vert^2 + \vert\beta\vert^2\big]\big[1-\cos(2B(t))\big]}\me^{-k^2\vert\eta(t)\vert^2\big[2\overbar{n} + 1\big]} \, , \label{eq:Duan_optical}
\end{split} 
\end{equation}


\noindent where $ \Bar{n}$ is the thermal occupation number for the mechanical oscillator.

\begin{figure}[t] 
    \centering
    \includegraphics[width=8.6 cm]{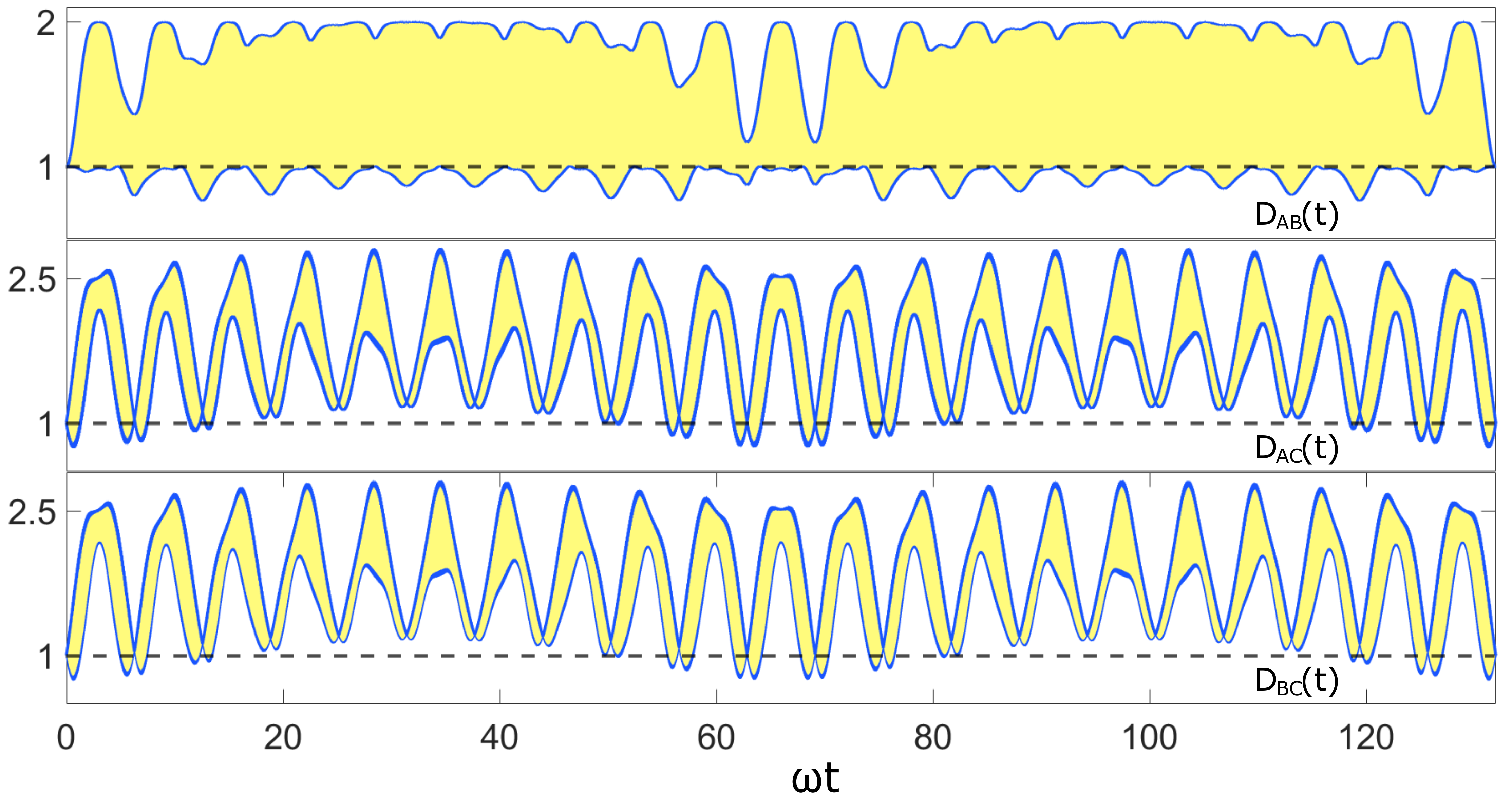}
    \caption[]{\small Time-dependent EPR variances $D_{ij}(t)$ (yellow) for the various bipartitions of the system, with its envelope (blue) and threshold for the Duan Criteria (black dashed line). We use $\omega_a = \omega_b = 10^{15}$ Hz and the remaining parameters as in Table~\ref{table:proposed_params}. 
    }
    \label{fig:Figure3}
\end{figure}

Figure \ref{fig:Figure3} shows the time evolving EPR variances for different bipartitions of the system: $ AB $ (opto-opto), $BC$ (opto-mechanical) and  $AC$ (opto-mechanical). We once again observe periodic birth of entanglement and, from the discussion in Section \ref{sec:Hamiltonian}, we can assert that there is death and revivals of entanglement for every bipartition of the system. Moreover, as in the qubit case, the appearance of entanglement between optical modes given the initially separable state can be used in experiments to probe the non-classicality of the mechanical mode \cite{Marletto2017}. 
Although the analytical formulas for the EPR variance are rather involved, it is possible to obtain insight into the entanglement dynamics by looking into their periodicity. 


For typical optical and mechanical frequencies, the term $\cos((r_a + r_b)t)$ represents fast oscillations that do not 
contribute significantly to the overall envelope of the EPR variances. Consequently, when $k \ll 1/\sqrt{2}$, equation \eqref{eq:Duan_optical} is dominated by the term $\exp{-2\big[\vert\alpha\vert^2 + \vert\beta\vert^2\big]\big[1-\cos(2B(t))\big]}$ which has a period of $\tau = \pi / k^2$. We call this the ``low coupling'' regime. 
On the other hand, if $k\gg 1/\sqrt{2}$ the variance is dominated by the term $\exp{-k^2\vert\eta(t)\vert^2\big[2\overbar{n} + 1\big]}$, of period $\tau = 2\pi$. We refer to this as the ``high coupling'' regime. 
The periodicity of these functions dictates the overall periodicity of the envelope of the EPR variances.
Going back to non-scaled time we make the substitution $\tau \rightarrow \tau/\omega_m$.  Then, for values in the low coupling regime, observation of EPR variance revivals are only possible when $ \pi / \omega_m k^2 \ll \kappa^{-1}$, where $\kappa^{-1}$ is the inverse cavity linewidth, or the approximate photon lifetime in the cavity. This translates into the so-called photon blockade condition $ g_{0}^{2} / (\omega_{m} \kappa) \gg 1 $ \cite{Neumeier2018, Rabl2011}.
For the high coupling regime, observation of full entanglement dynamics is conditioned on satisfying $2\pi / \omega_{m} \ll \kappa^{-1}$, which translates into the resolved-sideband regime $\omega_m \gg\kappa$ \cite{Meyer2019}.

\begin{figure}[t] 
    \centering
    \includegraphics[width=8.6 cm]{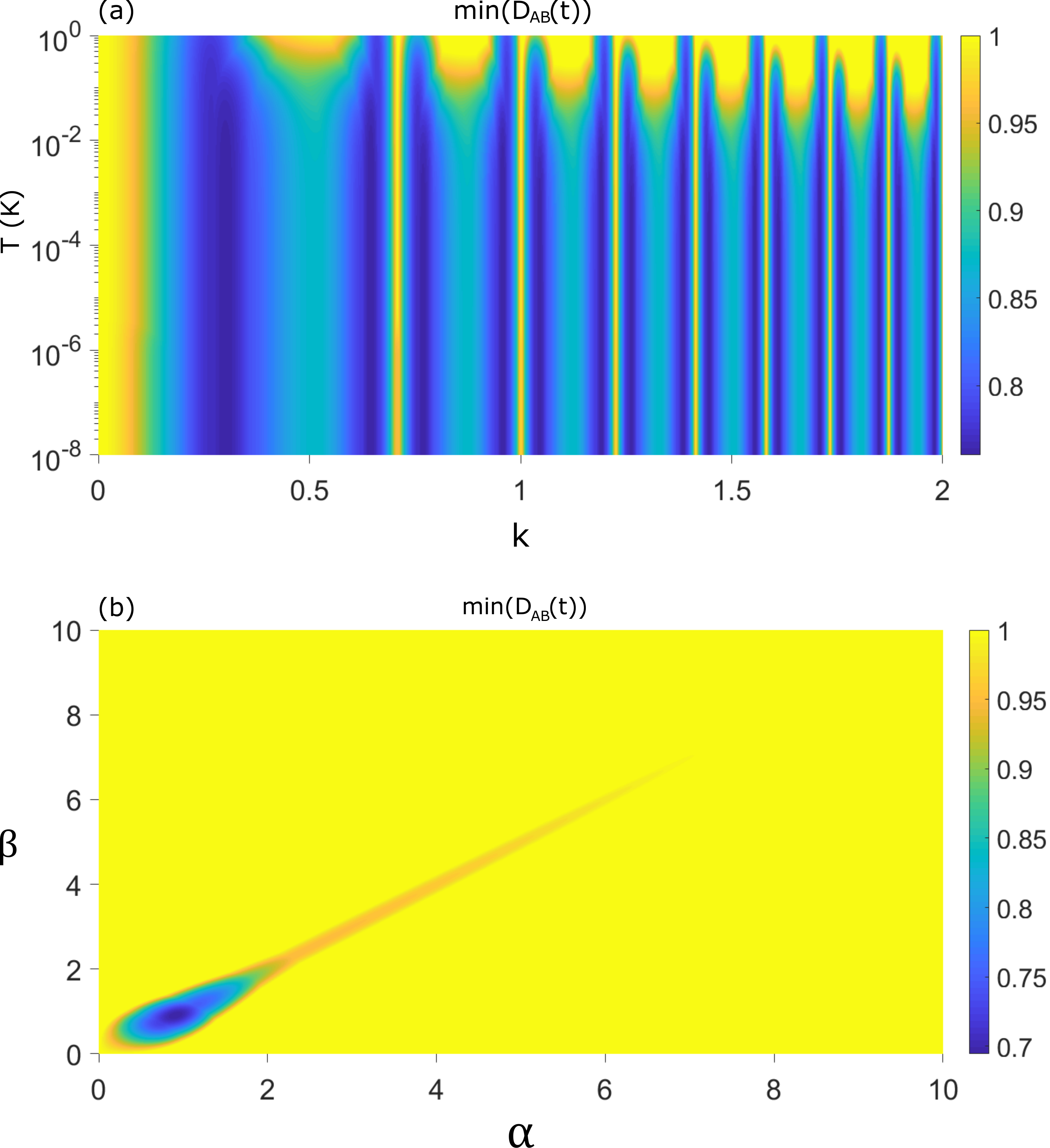}
    \caption[]{\small \textbf{(a)} Minimum value of $D_{AB}(t)$ within the photon lifetime inside the cavity, $\tau=\kappa^{-1} \sim 15.6\mu s$, as a function of \textbf{(a)} dimensionless coupling $k$ and mechanical oscillator's temperature $T$, and \textbf{(b)} coherent state amplitudes of $\alpha$ and $\beta$. The optimal coherent amplitudes that globally minimizes the Duan criteria are found to be $\alpha \sim \beta \sim 0.91$. We use $\omega_a = \omega_b = 10^{15}$ Hz and every other parameter as in in Table \ref{table:proposed_params}. 
    }
    \label{fig:Figure4}
\end{figure}

Knowledge of the behavior of the EPR variance as a function of optomechanical coupling and temperature is useful when designing an experiment. 
We observe that the optical entanglement in modes $ AB $ is strongly affected by the dimensionless coupling $ k $ in a non-trivial way. Figure \ref{fig:Figure4}(a) shows the minimum of the EPR variance for the optical modes within the photon lifetime $ \kappa^{-1}$ inside the cavity as a function of coupling $ k $ and oscillator initial temperature $ T $. 
In the low coupling regime, where $k \ll 1/\sqrt{2}$, above a threshold coupling of $ k \approx 0.1 $, $  \min(D_{AB}(t)) $ generally falls bellow one and hence the system always exhibits entanglement. 
In contrast, the high coupling regime, $k \gg 1/\sqrt{2}$, is highly sensitive to changes in $k$, presenting  local maxima at $2k^2 = N $, $ N $ a positive integer. For these maxima, 
no entanglement can be certified regardless of the oscillator temperature. For instance, if optical entanglement is to be maximized, increasing coupling may not be the best strategy.
Moreover we observe that entanglement persists well above the micro-kelvin temperatures reported in current optomechanical experiments \cite{Delic2019, Delic892, Windey2019, Murch2008}, consistent with the theoretical results for the linearized regime in \cite{Paternostro2007}. Conversely, from our numerical calculations, we have observed that although the minimum value of $D_{AB}(t)$ remains well below $1$, the time spent below this threshold becomes increasingly smaller as the temperature grows higher, making it effectively harder to experimentally verify entanglement with the Duan Criteria at micro-kelvin temperatures. 

The mean number of photons in the cavity also plays an important role in optical entanglement generation. Figure \ref{fig:Figure4}(b) shows a surface plot of $ \min(D_{AB}(t)) $ within the photon lifetime as a function of the coherent state amplitudes $ \alpha $ and $\beta$, taken to be real numbers for simplicity. The Duan Criteria can only be conclusive when the energy is approximately evenly distributed among the two optical modes, which happens when $\alpha\sim\beta$. For the parameters used, the optimal coherent amplitudes that minimizes $D_{AB}(t)$ are found to be $\alpha \sim \beta \sim 0.91$. 

\section{Experimental proposal}
With increasing advances in the field of quantum cavity optomechanics \cite{Delic892,Delic2019, Windey2019}, experiments in the high coupling and long coherence time regimes are expected, although observing entanglement as described in the present work remains challenging. 
One notable exception and a promising candidate is optomechanics with ultracold atomic ensembles, where a coherent cloud of atoms is trapped within an optical cavity and the collective center of mass coordinate effectively behaves as a quantum mechanical oscillator. Couplings as high as $ k \approx 10$ have been reported in such ultracold experiments \cite{Murch2008}, and the system allows wide tunability of the relevant parameters. 

\begin{table}[h!]
\caption{Proposed values for the experimental implementation with ultracold atoms. \label{tab1}}
\centering
\resizebox{\columnwidth}{!}{
\begin{tabular}{lcc}
\hline
\hline
Parameter  \hspace{2em}  &          Units     \hspace{2em}      & Value         \\ \hline
Number of atoms $N$    &       -        & $5.43\times10^5$ \\
Trap frequency $\omega_m$   &  kHz & 600        \\
Coupling \ $k = \frac{g_{0} }{ \omega_{m}}$ &   -   & 0.74
\\
Mechanical dissipation \ $\Gamma / 2\pi$ &   kHz   & 1
\\
Cavity Finesse $ \mathcal{F} $    &      -         & $3\times10^6$ \\
Cavity Length $ L $      &          $\mu$m     & 783          \\
Cavity Linewidth $ \kappa $      &          kHz     & 64\\
Temperature $ T $         &       $\mu$K        & 0.8       \\
Mean photon number $\vert \alpha \vert^{2}= \vert \beta \vert^{2}$ & - & 0.25 \\
\hline\hline
\end{tabular}
\label{table:proposed_params}
}
\end{table}


In order to observe entanglement dynamics, the lifetime of a photon inside the cavity 1/$\kappa$ needs to be longer than the time necessary for optical entanglement to reach its first local maximum. 
Table \ref{tab1} presents values for the relevant experimental parameters adapted from \cite{Murch2008}, where the motion of an ultracold gas of ${}^{87} $Rb was studied. 
With these values the resulting photon lifetime is $\tau_p = 15.7\times10^{-6}$ s and within this lifetime the minimum value for the optical EPR variance is found to be $ D_{AB}(t)  \approx 0.8 $. This suggests that in these systems the observation of 
mechanically-induced optical entanglement can be within reach.

Moreover, following \cite{Murch2008}, we estimate the maximal heating rate of a cloud of atoms due to backaction from each optical mode, $R_{\rm c} = Ng_0^2/(4\Gamma \kappa) R_{\rm fs}$, and due to spontaneous emission, $R_{\rm fs} = (\hbar k_p)^2/m g_0^2 \overbar{n}_{\rm cav} \Gamma/ \Delta_{\rm ca}$, where $k_p$ is the wave vector of the optical modes, $m$ is the atomic mass of ${}^{87} $Rb,  $\overbar{n}_{\rm cav}$ is the mean photon number for each optical mode and $\Delta_{\rm ca}$ is the detuning between the frequencies for the optical modes and the atomic resonance frequency. Given the proposed parameters, heating is dominated by $R_{\rm c} \gg R_{\rm fs}$. We note, that during the lifetime of a single photon inside the cavity, we estimate the ratio of the maximal energy delivered to the atomic cloud and its initial thermal energy to be $ (R_{\rm c} \tau_p)/(K_B T)\sim 1.6\cdot 10^{-16}$. Therefore, the heating of the atomic ensemble should be negligible during the course of the experiment.




\section{Conclusion}



In this article we have studied the entanglement and entropy dynamics of a ``mirror-in-the-middle" optomechanical system. Implementations using levitated particles have been briefly discussed.
We have seen that an initially separable quantum state can evolve to an entangled one, exhibiting birth, death and revivals of entanglement and entropy for qubit and continuous variable states; moreover,  the appearance of entanglement in this setting evidences the non-classical nature of the mechanical oscillator.  Therefore, optical entanglement will arise if and only if the mechanical oscillator is quantum mechanical.

The entanglement dynamics is strongly influenced by the system parameters, notably the dimensionless optomechanical coupling $ k = g_{0} / \omega_{m} $ and the mean number of photons in the experiment. 
We have shown the existence of two distinct regimes depending on whether $ k < 1/\sqrt{2} $ or $ k \geq 1/\sqrt{2} $. In addition, we have observed that optical entanglement is maximized when the energy is evenly distributed in the optical modes and that it persists at micro-kelvin temperatures of the mechanical mode. These are valuable informations when designing an experiment.
Optomechanics with ultracold atomic ensembles presents an interesting candidate for implementing the studied entanglement dynamics. 
Although a promising candidate, the dispersive Hamiltonian is not the only available platform to untangle the dynamics of entanglement and information flow in optomechanical systems. Exploring alternatives such as coherent scattering \cite{Delic2019, Windey2019, RiosSommer2020, Hornberger2020_CS,Radim2020,Chauhan2020} might prove to be a very fruitful approach to observe entanglement and non-classicality in experimental optomechanical systems.

\section{Acknowledgments} This work was financed in part by the Serrapilheira Institute (grant number Serra-1709-21072), by Coordenac\~ao de Aperfei\c{c}oamento de Pessoal de N\'ivel Superior - Brasil (CAPES) - Finance Code 001 and by Conselho Nacional de Desenvolvimento Cient\'ifico e Tecnol\'ogico (CNPq). I. B. thanks the support received by the FAPERJ Scholarship No. E-26/200.270/2020. T.G. thanks the  support received by the FAPERJ Scholarship No. E-26/202.830/2019.

\appendix
\section{Implementation using levitated nanoparticles} \label{appendix:levitated_implementation}
Consider the system shown in Figure \ref{fig:Figure1}(b): a nanoparticle of radius $r$, mass $m$ and refractive index $n_p$ is trapped in an harmonic trap of frequencies $\omega_{j=x,y,z}$ created by an optical tweezer inside a cavity populated by two optical modes of frequencies $\omega_{a}$ and $\omega_b$ and annihilation/creation operators $\hat{a}$/$\hat{a}^\dagger$ and $\hat{b}$/$\hat{b}^\dagger$, respectively. The presence of the particle causes a position dependent shift on the cavity's resonance frequencies, so that the Hamiltonian of this system becomes \cite{Kiesel2013}
\begin{eqnarray}
\frac{\hat{H}}{\hbar}= \omega_a\hat{a}^\dagger \hat{a}-U_{0,a} \sin^2[k_a(x_0+x)]\hat{a}^\dagger \hat{a}+\omega_b\hat{b}^\dagger \hat{b} \nonumber \\ -U_{0,b} \sin^2[k_b(x_0+x)]\hat{b}^\dagger \hat{b}+\sum_{j=x,y,z}\omega_{j} \hat{c}_j^\dagger \hat{c}_j \, ,
\label{eq:levitated_hamiltonian}
\end{eqnarray}
where $x_0$ is the position of the center of the trap, $x$ is the particle's displacement, $U_{0,i}=\omega\hat{a}_i\alpha/2\epsilon_0V_i$ is the frequency shift when the particle is at an intensity maximum of the cavity, $\alpha=4\pi\epsilon_0r^3(n_p^2-1)/(n_p^2+2)$ is the polarizability of the particle, $V_i$ and $k_i$ are the volume and the wavenumber of mode $i$, respectively, and $\hat{c}_j$/$\hat{c}_j^\dagger$ is the phonon annihilation/creation operators along axis $j=x,y,z$. We will not consider driving terms since we are interested in the dynamics of an isolated cavity without the influence of external systems such as a driving laser, as explained in the main text. 


The interaction terms may yield linear couplings between the optical modes and the sphere if $x_0$, $k_a$ and $k_b$ are properly chosen. Expanding $\sin^2{k_i(x_0+x)}$ around $x_0$ gives
\begin{equation}
\begin{split}
     \sin^2&[k_i(x_0+x)]=\sin^2(k_ix_0) \\ &+ k_a\sin(2k_ax_0)x+ 2k_a^2\cos(2k_ax_0)x^2+\mathcal{O}(x^3).
\end{split}
\end{equation}

Now, consider the particular case in which the frequencies of the optical modes are two consecutive resonance frequencies, and let $L=2n(\lambda_a/2)=(2n+1)(\lambda_b/2)$ without loss of generality. Then, if the sphere is placed near the center of the cavity at $x_0=L/2+\lambda_a/8\approx L/2+\lambda_b/8$, we have
\begin{equation}
    \sin^2{k_a(x_0+x)}\approx 1/2+k_a x
\end{equation}
and 
\begin{equation}
    \sin^2{k_b(x_0+x)}\approx 1/2-k_b x.
\end{equation}

Substituting these approximations in equation \eqref{eq:levitated_hamiltonian} and disregarding the motion along the $y$ and $z$ axes, we get
\begin{equation}
\label{eq:levitated_hamiltonian_approximation}
\hspace*{-1mm}\frac{\hat{H}}{\hbar}= \omega_a'\hat{a}^\dagger \hat{a}\nonumber-g_{0}\hat{a}^\dagger \hat{a}(\hat{c}_x^\dagger+\hat{c}_x)+\omega_b'\hat{b}^\dagger \hat{b}+g_{0}\hat{b}^\dagger \hat{b}(\hat{c}_x^\dagger+\hat{c}_x)+\omega_{x} \hat{c}_x^\dagger \hat{c}_x \, ,
\end{equation}
where $\omega_{a}'=\omega_{a}-U_{0,i}/2$ and  $g_0=U_{0,a}k_ax_{ZPF}\approx U_{0,b}k_bx_{ZPF}$ is the intended linear coupling, with $x_{ZPF}=\sqrt{\hbar/2m\omega_x}$ the zero-point fluctuation. This Hamiltonian is formally equivalent to that of a double-sided cavity with a moving ``mirror-in-the-middle'' analysed in the main text, with the condition $g_{0,a}\approx g_{0,b}$ being met due to the use of consecutive resonance frequencies. As a final remark, note that the modes $A$ and $B$ are not used for cooling of the particle, which is necessary for observing optical entanglement in the system \cite{Marletto2017}. Cooling of the center-of-mass motion of the particle can be addressed in a number of ways, such as a feedback acting on the trapping laser \cite{Gieseler2012, Conangla2019}, dispersive coupling with a third optical mode \cite{Meyer2019} or coherent scattering between the trapping beam and the optical cavity \cite{Windey2019, Delic2019, RiosSommer2020}.

\section{Implementation using ultracold atomic ensembles} \label{appendix:ultracol_atom_implementation}

In ultracold atom optomechanical experiments, an atomic ensemble is trapped inside an optical cavity. Collective center-of-mass motion of the atoms alters the cavity resonance frequency. It is similar to the dispersive optomechanical experiments with levitated spheres described in the previous section, with the cloud of atoms playing the role of levitated nanoparticle. 

The optomechanical coupling between an ultracold atomic cloud and a cavity optical mode with wavelength $\lambda_a$ is \cite{Murch2008}
\begin{equation}
        g_{0,a} = k_{a} N \dfrac{\alpha_{0}^{2}}{\Delta_{ca}}  \sin ( 2k_{a} z_{0} ) \sqrt{\dfrac{\hbar}{2 N m \omega_{m}}} \, ,
\end{equation}
where $k_a$ is the wavenumber, $N$ is the number of atoms, $\Delta_{ca}$ is the atom-cavity detuning, $m$ is the mass of a single atom, $\omega_m$ is the mechanical frequency and $\alpha_{0} = \sqrt{d^{2} \omega_{c}/2 \hbar \epsilon_{0} V_{c}} $ is the atom-single photon coupling rate, with $d$ the dipole moment for the transition between the relevant atomic levels and $V_c$ the cavity volume. Reported values from Ref. \cite{Murch2008} for these quantities are shown in Table \ref{tab:Murch}. As in the previous case of a levitated nanoparticle, we chose $\sin{(2k_az_0)}=1$. The wavelength $\lambda_b$ should be chosen so that $\sin{(2k_bz_0)}=-1$, as to provide $g_{0,a}=g_{0,b}$.

\begin{table}[h!]
\caption{Values reported in \cite{Murch2008}. \label{tab:Murch}}
\centering
\begin{tabular}{lcc}
\hline
\hline
Parameter  \hspace{2em}  &          Units     \hspace{2em}      & Value         \\ \hline
Number of atoms    &       -        & $10^5$ \\
Trap frequency $\omega_m$   &  kHz & 2$\pi\times$40        \\
Coupling \ $k = \frac{g_{0} }{ \omega_{m}}$ &   -   & 9.50      \\
Cavity Finesse $ \mathcal{F} $    &      -         & $5.8\times10^5$ \\
Cavity Length $ L $      &          $\mu$m     & 194          \\
Cavity mirror's radius $ R $      &          cm     & 5          \\
Cavity Linewidth $ \kappa $      &          MHz     & 2$\pi\times$0.66          \\
Temperature $ T $         &       $\mu$K        & 0.8        \\
\hline\hline
\end{tabular}
\end{table}

As discussed in the main text, for the entanglement dynamics experiment to be feasible, the photon-lifetime $\tau_p$ must be greater than the entanglement period $\tau_e$. One of the main constraints in fulfilling this condition is set by the Finesse of the cavity. Therefore, we look for the minimum value of $\tau_e/\tau_p$ by varying the cavity length $L$, the number of atoms $N$ and the mechanical frequency $\omega_m$ around the values in Table \ref{tab:Murch}, and then calculate the minimum Finesse necessary to make $\tau_e/\tau_p<1$ given the optimal values of $L$, $N$ and $\omega_m$.

In doing these calculations, it is necessary to account for the changes in the mode volume, given by $V_c=\pi w_a^2 L$, where 
\begin{equation}
    w_a=\sqrt{\frac{\lambda_a}{2\pi}\sqrt{L(2R-L)}}
\end{equation}
is the mode's waist, and the changes in the cavity linewidth
\begin{equation}
    \kappa= \frac{\mathcal{F}}{\nu_{FSR}}
\end{equation}
with $\nu_{FSR}=c/2L$ the cavity's free spectral range. Finally, it is important to make sure that $k=g_0/\omega_m\neq\sqrt{n/2}$, $n$ a positive integer, when the Duan Criteria is inconclusive.
We find that for $L=783 \ \mu$m, $N=5.43\times10^5$ and $\omega_m=2 \pi \times 95$ kHz the dimensionless optomechanical coupling is $k=0.743$ and the entanglement period to photon lifetime ratio is $\tau_e/\tau_p=3.46$. The Finesse should then be increased to $2.01\times10^6$, so that $\tau_e/\tau_p \simeq 1$ and entanglement becomes measurable. In the main text, the proposed Finesse is about $1.5$ times larger, so that $\tau_e/\tau_p  = 0.669$.

Another parameter that could be varied are the radii of the cavity mirrors. Considering $1$ cm, $2.5$ cm, $5$ cm and $10$ cm as possible radii, we find the values presented in Table \ref{tab:final_exp}. As we can see, the Finesse constraint can be relaxed provided that a smaller radius is used. Overall, the necessary values for $N$, $\omega_m$ and $\mathcal{F}$ differ by less than one order of magnitude from reported values in the literature. 
\hspace*{-1em}
\begin{table}[h!]
\caption{Optimal parameters for different cavity mirror's radii. \label{tab:final_exp}}
\centering
\resizebox{\columnwidth}{!}{
\begin{tabular}{ccccc}
\hline
\hline
\hspace{1em}R(cm)  \hspace{1em}  & \hspace{1em} $L$($\mu$m)    \hspace{1em}      & \hspace{1em}$N$($10^5$) \hspace{1em}     & \hspace{1em}$\omega_m$(kHz) \hspace{1em}  & \hspace{1em}$\mathcal{F}$($10^6$)\hspace{1em}     \\ \hline
1   &  1211 & 3.85 &        2$\pi\times$95  &  1.30\\
2.5   & 1035  & 5.64    & 2$\pi\times$92&  1.57\\
5 &   783   & 5.43  & 2$\pi\times$95  &  2.01\\
10    &     669         & $5.80$ & 2$\pi\times$91& 2.47
\\
\hline\hline
\end{tabular}
}
\end{table}

\section{The unitary evolution} \label{appendix:unitary_evolution}
The ``mirror-in-the-middle'' Hamiltonian can be written as
\begin{equation}
\frac{\hat{H}}{\hbar} = \omega_m \hat{c}^{\dagger}\hat{c} + \omega_a \hat{a}^{\dagger}\hat{a} + \omega_b \hat{b}^{\dagger}\hat{b} - g_0\left(\hat{a}^{\dagger}\hat{a} - \hat{b}^{\dagger}\hat{b}\right)(\hat{c}+\hat{c}^{\dagger}),
\end{equation}
where $\hat{a}$, $\hat{b}$ and $\hat{c}$ denote annihilation operators for the optical mode in the left cavity ($A$), right cavity ($B$) and the mechanical oscillator ($C$), respectively. In what follows we will work with the re-scaled time $\omega_m t$, denoted henceforth by $t$, and introduce the dimensionless variables $r_a= \omega_a/\omega_m$, $r_b= \omega_b/\omega_m$ and $k=g_0/\omega_m$.

Define the unitary operator \cite{Bose1997}
\begin{equation}
   \hat{E}(k) = \exp(k (\hat{a}^{\dagger}\hat{a} - \hat{b}^{\dagger}\hat{b})(\hat{c}^{\dagger}-\hat{c})).
\end{equation}
The operator $\hat{E}(k)$ commutes with $\hat{a}^{\dagger}\hat{a}$ and $\hat{b}^{\dagger}\hat{b}$, but not with $\hat{c}$,
\begin{equation}
\label{eq:EcE}
    \hat{E}(k)^{\dagger} \hat{c} \hat{E}(k) = \hat{c} + k(\hat{a}^{\dagger}\hat{a} - \hat{b}^{\dagger}\hat{b}),
\end{equation}
which is calculated through the general identity,
\begin{equation}
\label{General Identity}
e^{-\hat{A}} \hat{B} e^{+\hat{A}} = \hat{B} + \comm{\hat{B}}{\hat{A}} + \frac{1}{2!} \comm{\comm{\hat{B}}{\hat{A}}}{\hat{A}} + \cdots\ \ .
\end{equation}

Using equation \eqref{eq:EcE} and its adjoint, we have
\begin{equation}
\label{EHE}
    \hat{E}(k)^{\dagger} \frac{\hat{H}}{\hbar\omega_m} \hat{E}(k) = \hat{c}^{\dagger}\hat{c} + r_a \hat{a}^{\dagger}\hat{a} + r_b \hat{b}^{\dagger}\hat{b} - k^2(\hat{a}^{\dagger}\hat{a}-\hat{b}^{\dagger}\hat{b})^2.
\end{equation}
Considering the number basis $\lbrace \ket{n,m,\ell} \rbrace$, we see by the equation above that $\hat{E}(k)\ket{n,m,\ell}$ are the energy eigenstates of the system. Those are of the form:
\begin{equation}
\hat{E}(k)\ket{n,m,\ell} = \hat{\mathcal{D}}_C(k(n-m))\ket{n,m,\ell},
\end{equation}
where we denote the displacement operator of the mechanical oscillator by $\hat{\mathcal{D}}_C(\kappa)=\exp(\kappa \hat{c}^{\dagger} - \kappa^* \hat{c})$, where $\kappa$ is a complex number. The energies corresponding to the eigenstates above are
\begin{equation}
E_{n,m,\ell}=\hbar\omega_m\ell + \hbar\omega_a n + \hbar\omega_b m - \hbar \omega_m k^2(n-m)^2.
\end{equation}

By exponentiation of equation $(\ref{EHE})$, the unitary evolution operator is found to be
\begin{equation}
    \hspace*{-2mm}\hat{U}(t) = \hat{E}(k) e^{-i \left(\hat{c}^{\dagger}\hat{c} + r_a \hat{a}^{\dagger}\hat{a} + r_b \hat{b}^{\dagger}\hat{b} - k^2(\hat{a}^{\dagger}\hat{a}-\hat{b}^{\dagger}\hat{b})^2\right) t} \hat{E}(k)^{\dagger}.
    \label{evolution_unitary}
\end{equation}
Next, we recall that $e^{i \hat{c}^{\dagger}\hat{c} t} \hat{c}^{\dagger} e^{-i \hat{c}^{\dagger}\hat{c} t} = e^{it} \hat{c}^{\dagger}$ and $e^{i \hat{c}^{\dagger}\hat{c} t} \hat{c} e^{-i \hat{c}^{\dagger}\hat{c} t} = e^{-it} \hat{c}$, from which we derive the following identity
\begin{equation}
\begin{split}
    e^{i \hat{c}^{\dagger}\hat{c} t} \hat{A} (\hat{c}^{\dagger} - \hat{c})e^{-i \hat{c}^{\dagger}\hat{c} t} = \hat{A} (e^{it} \hat{c}^{\dagger} - e^{-it}\hat{c}) \\ \Rightarrow e^{i \hat{c}^{\dagger}\hat{c} t}e^{\hat{A} (\hat{c}^{\dagger} - \hat{c})} e^{-i \hat{c}^{\dagger}\hat{c} t} = e^{\hat{A} (e^{it} \hat{c}^{\dagger} - e^{-it}\hat{c})} ,
\end{split}
\end{equation}
where $\hat{A}$ is any operator that commutes with both $\hat{c}$ and $\hat{c}^{\dagger}$.

Letting $\hat{A} = k (\hat{a}^{\dagger}\hat{a} - \hat{b}^{\dagger}\hat{b})$ in the identity above, we conclude that
\begin{equation}
\begin{split}
     \hat{E}(k) &e^{-i \hat{c}^{\dagger}\hat{c} t} = e^{-i \hat{c}^{\dagger}\hat{c} t} e^{i \hat{c}^{\dagger}\hat{c} t} \hat{E}(k) e^{-i \hat{c}^{\dagger}\hat{c} t} \\ &= e^{ - i \hat{c}^{\dagger}\hat{c} t} \exp(k (\hat{a}^{\dagger}\hat{a} - \hat{b}^{\dagger}\hat{b})(\hat{c}^{\dagger} e^{it}-\hat{c} e^{-it})) ,
\end{split}
\end{equation}
hence
\begin{equation}
\begin{split}
    \label{Product of Exponentials}
    \hat{E}(k)& e^{-i \hat{c}^{\dagger}\hat{c} t} \hat{E}(k)^{\dagger} = \\ &e^{ - i \hat{c}^{\dagger}\hat{c} t}\times \exp(k (\hat{a}^{\dagger}\hat{a} - \hat{b}^{\dagger}\hat{b})(\hat{c}^{\dagger} e^{it}-\hat{c} e^{-it})) \\ &\times \exp(-k (\hat{a}^{\dagger}\hat{a} - \hat{b}^{\dagger}\hat{b})(\hat{c}^{\dagger}-\hat{c})) .
\end{split}
\end{equation}
From the commutator
\begin{equation}
    \frac{1}{2} [\hat{c}^{\dagger}e^{it} - \hat{c} e^{-it}, \hat{c}^{\dagger} - \hat{c}] = i\sin(t),
\end{equation}
it becomes straightforward to compute
\begin{equation}
\begin{split}
     [k (\hat{a}^{\dagger}\hat{a} - \hat{b}^{\dagger}\hat{b})(\hat{c}^{\dagger} e^{it}-\hat{c} e^{-it}), k (\hat{a}^{\dagger}\hat{a} - \hat{b}^{\dagger}\hat{b})(\hat{c}^{\dagger}-\hat{c})] \\ = 2ik^2(\hat{a}^{\dagger}\hat{a} - \hat{b}^{\dagger}\hat{b})^2 \sin(t) .
\end{split}
\end{equation}
Since this commutes with both $(\hat{a}^{\dagger}\hat{a} - \hat{b}^{\dagger}\hat{b})(\hat{c}^{\dagger} e^{it}-\hat{c} e^{-it})$ and $(\hat{a}^{\dagger}\hat{a} - \hat{b}^{\dagger}\hat{b})(\hat{c}^{\dagger}-\hat{c})$, equation ($\ref{Product of Exponentials}$) can be simplified using the particular form of the Baker-Campbell-Haursdoff formula 
\begin{equation}
    e^{\hat{A}} e^{\hat{B}} = e^{\hat{A}+\hat{B} + \frac{1}{2} [\hat{A},\hat{B}]} ,
\end{equation}
valid when operators $\hat{A}$ and $\hat{B}$ commute with $[\hat{A},\hat{B}]$. Thus, the expression 
\begin{equation}
\begin{split}
    \label{Evolution_Operator}
    \hat{U}(t) &= e^{-i \hat{c}^{\dagger}\hat{c} t} e^{-i r_a \hat{a}^{\dagger}\hat{a} t} e^{-i r_b \hat{b}^{\dagger}\hat{b} t}  \\ &\times e^{k(\hat{a}^{\dagger}\hat{a}-\hat{b}^{\dagger}\hat{b})(\hat{c}\eta(t)-\hat{c}^{\dagger}\eta(t)^*)} e^{-i(\hat{a}^{\dagger}\hat{a}-\hat{b}^{\dagger}\hat{b})^2 B(t)} 
\end{split}
\end{equation}
holds for the unitary evolution operator $\hat{U}(t)$, where $\eta(t) = 1 - e^{-it}$ and $B(t)= -k^2 (t-\sin(t))$. 

In order to arrive at the expression for $\hat{U}(t)$ exactly as presented in the main text, we move the term $e^{-i \hat{c}^{\dagger}\hat{c} t}$ to the right, which gives 
\begin{equation}
\begin{split}
        &\hat{U}(t) =  e^{-i r_a \hat{a}^{\dagger}\hat{a} t} e^{-i r_b \hat{b}^{\dagger}\hat{b} t} \\ &\times e^{k(\hat{a}^{\dagger}\hat{a}-\hat{b}^{\dagger}\hat{b})(\eta(t)\hat{c}^{\dagger}-\eta(t)^*\hat{c})} e^{-i(\hat{a}^{\dagger}\hat{a}-\hat{b}^{\dagger}\hat{b})^2 B(t)}  e^{-i \hat{c}^{\dagger}\hat{c} t} \, .
\end{split}
\end{equation}

In the interaction picture, we evolve states according to the unitary operator
\begin{equation}
\hat{U}_{\text{I.P.}}(t) = e^{k(\hat{a}^{\dagger}\hat{a}-\hat{b}^{\dagger}\hat{b})(\hat{c}\eta(t)-\hat{c}^{\dagger}\eta(t)^*)} e^{-i(\hat{a}^{\dagger}\hat{a}-\hat{b}^{\dagger}\hat{b})^2 B(t)}.
\end{equation}

The Heisenberg evolution of the annihilation operators $\hat{a}$, $\hat{b}$ and $\hat{c}$ can then be derived,
\begin{equation}
\begin{split}
\label{Heisenberg_a}
    &\hat{a}(t) = U^{\dagger}(t) \hat{a} \hat{U}(t) \\ &= e^{-i(r_a t + B(t))} e^{2i B(t) \hat{b}^{\dagger}\hat{b}} e^{-2i B(t) \hat{a}^{\dagger}\hat{a}} \hat{\mathcal{D}}_C (+k \xi(t)) \hat{a},
\end{split}
\end{equation}
\begin{equation}
\label{Heisenberg_b}
\begin{split}
    &\hat{b}(t)= U^{\dagger}(t) \hat{b} \hat{U}(t) \\ &= e^{-i(r_b t + B(t))} e^{-2i B(t) \hat{b}^{\dagger}\hat{b}}e^{2i B(t) \hat{a}^{\dagger}\hat{a}} \hat{\mathcal{D}}_C ( - k \xi(t)) \hat{b},
\end{split}
\end{equation}
\begin{equation}
\label{Heisenberg_c}
\hat{c}(t) =  U^{\dagger}(t) \hat{c} \hat{U}(t) = \hat{c} e^{-it} + k (\hat{a}^\dagger \hat{a} - \hat{b}^\dagger \hat{b})\eta(t),
\end{equation}
where $\xi(t)=e^{it}\eta(t)= e^{it} - 1 = -\eta(t)^*$. In deriving these expressions we made use of the relations
\begin{equation}
\label{Trick 1}
    e^{i \hat{A} \hat{a}^{\dagger}\hat{a}} \hat{a} e^{-i\hat{A}\hat{a}^{\dagger}\hat{a}} = e^{-i \hat{A}}\hat{a} ,
\end{equation}
\begin{equation}
\label{Trick 2}
    e^{i \hat{A} (\hat{a}^{\dagger}\hat{a})^2} \hat{a} e^{-i\hat{A}(\hat{a}^{\dagger}\hat{a})^2} = e^{-i \hat{A}(2\hat{a}^{\dagger}\hat{a} + I)}\hat{a} ,
\end{equation}
applicable whenever $\hat{A}$ commutes with both $\hat{a}$ and $\hat{a}^\dagger$. For instance, equation ($\ref{Heisenberg_c}$) is derived as
\begin{equation}
\begin{split}
    U^{\dagger}(t) \hat{c} \hat{U}(t) &= e^{-k(\hat{a}^{\dagger}\hat{a}-\hat{b}^{\dagger}\hat{b})(\hat{c}\eta(t)-\hat{c}^{\dagger}\eta(t)^*)} \\ &\times  e^{+i \hat{c}^{\dagger}\hat{c} t}\hat{c} e^{-i \hat{c}^{\dagger}\hat{c} t} e^{k(\hat{a}^{\dagger}\hat{a}-\hat{b}^{\dagger}\hat{b})(\hat{c}\eta(t)-\hat{c}^{\dagger}\eta(t)^*)}  \\ &= e^{-it} e^{-k(\hat{a}^{\dagger}\hat{a}-\hat{b}^{\dagger}\hat{b})(\hat{c}\eta(t)-\hat{c}^{\dagger}\eta(t)^*)}  \\ &\times\hat{c} e^{k(\hat{a}^{\dagger}\hat{a}-\hat{b}^{\dagger}\hat{b})(\hat{c}\eta(t)-\hat{c}^{\dagger}\eta(t)^*)} \\  &= \hat{c} e^{-it} - k\eta(t)^* e^{-it} (\hat{a}^{\dagger}\hat{a}-\hat{b}^{\dagger}\hat{b}),
\end{split}
\end{equation}
which leads to the desired result after noting that $-\eta(t)^* e^{-it} = \eta(t)$.

\section{ Duan Criteria} \label{appendix:Duan_Criteria}
A general sufficient condition for non-separability of a bipartite continuous variable system is provided by Duan criteria $\cite{DuanCriteria}$. For any bipartite quantum system with observables $\hat{X}_1,\hat{P}_1,\hat{X}_2,\hat{P}_2$ satisfying the canonical commutation relations
\begin{equation}
    [\hat{X}_j,\hat{P}_k] = i\delta_{jk},
\end{equation}
separability of the state $\rho$ implies that
\begin{equation}
    D = \frac{\big(\Delta(\hat{X}_1+\hat{X}_2)\big)^2}{2} + \frac{\big(\Delta(\hat{P}_1-\hat{P}_2)\big)^2}{2}\geq 1.
\end{equation}
In other words, if we verify that $D<1$, then we may conclude that the modes are in an entangled state. We stress that $D\geq 1$ provides no information: this inequality does not imply that the state is separable.

If the operators above are expressed in terms of annihilation and creation operators, defined by
\begin{equation}
    \hat{X}_j = \frac{1}{\sqrt{2}} (\hat{a}_j^{\dagger}+\hat{a}_j) \ \ , \ \ \hat{P}_j = \frac{i}{\sqrt{2}} (\hat{a}_j^{\dagger} - \hat{a}_j),
\end{equation}
then the following expression is useful:
\begin{align}
    D = \langle \hat{a}_1^{ \dagger}\hat{a}_1 \rangle &+ \langle \hat{a}_2^{ \dagger}\hat{a}_2 \rangle + \langle\hat{a}_1^{ \dagger}\hat{a}_2^{ \dagger}\rangle + \langle\hat{a}_1\hat{a}_2\rangle \nonumber \\
    &- \bigg( \langle\hat{a}_1^{ \dagger}\rangle + \langle\hat{a}_2\rangle \bigg) \bigg( \langle\hat{a}_2^{ \dagger}\rangle + \langle\hat{a}_1\rangle \bigg) + 1\, .
\end{align}
By simple algebraic manipulations, we find
\begin{equation}
\begin{split}
        D &= \langle \hat{a}_1^\dagger \hat{a}_1\rangle + \langle \hat{a}_2^\dagger \hat{a}_2\rangle + 2\Re\left(\langle \hat{a}_1\hat{a}_2\rangle - \langle \hat{a}_1\rangle \langle \hat{a}_2\rangle \right) \\ &- |\langle \hat{a}_1\rangle|^2 - |\langle \hat{a}_2\rangle|^2 + 1.
\end{split}
\end{equation}

For the optical modes, $\hat{a}_1\to \hat{a}$ and $\hat{a}_2\to\hat{b}$, this gives
\begin{equation}
   \begin{split}
    D_{AB}(t) = 1 + \bigg[|\alpha|^2 +  2\alpha\beta \cos\big((r_a+r_b)t\big) + |\beta|^2 \bigg] \\ - \bigg[|\alpha|^2 + 2\alpha\beta\cos\big( (r_a+r_b)t + 2B(t) \big) + |\beta|^2\bigg]\\ \times  \me^{-2\big[\vert\alpha\vert^2 + \vert\beta\vert^2\big]\big[1-\cos(2B(t))\big]}\me^{-k^2\vert\eta(t)\vert^2\big[2\overbar{n} + 1\big]} \, ,
\end{split} 
\end{equation}
assuming $\alpha,\beta\in\mathbb{R}$ for simplicity. The corresponding quantity between an optical mode and a mechanical mode, say $\hat{a_1}\to\hat{a}$ and $\hat{a_2}\to\hat{c}$, is given by
    \begin{align}
        &\;D_{AC}(t) = 1 + |\alpha|^2 + \overline{n} +2\Re(R(\alpha, t)) \\
        &+ \vert\eta(t)\vert^2\bigg[ k^2\big[  \vert\alpha\vert^2 + \vert\beta\vert^2 + (\vert\alpha\vert^2 - \vert\beta\vert^2)^2 \big] - k\big[\vert\alpha\vert^2-\vert\beta\vert^2\big]\bigg] \nonumber \\
        &- |\alpha|^2 e^{-2[|\alpha|^2+|\beta|^2](1-\cos(2B(t))} e^{-k^2\vert\eta(t)\vert^2[2\overline{n}+1]}\nonumber \, ,
    \end{align}

\noindent where
\begin{align}
    R&(\alpha, t)  \equiv \\ &\;\,\alpha e^{-\vert\alpha\vert^2(1-e^{-2iB(t)})-\vert\beta\vert^2(1-e^{2iB(t)})} \nonumber \\ &\times \left(\overline{n}+1-|\alpha|^2(1-e^{-2iB(t)}) + |\beta|^2(1-e^{2iB(t)})\right) \nonumber \\ & \times k\eta(t) e^{-i(r_a t + B(t))} e^{-k^2|\eta(t)|^2/2} e^{-\overline{n}k^2|\eta(t)|^2} \nonumber \, .
\end{align}

\bibliographystyle{unsrt}
\bibliography{main.bib}

\begin{thebibliography}{10}

\bibitem{Schroedinger1935}
E.~Schrodinger.
\newblock Die gegenwärtige situation in der quantenmechanik.
\newblock {\em Sci. Nat.}, 23(48):807--812, nov 1935.

\bibitem{Vedral2008}
V.~Vedral.
\newblock Quantifying entanglement in macroscopic systems.
\newblock {\em Nature}, 453(7198):1004--1007, jun 2008.

\bibitem{Preskill2018}
J.~Preskill.
\newblock Quantum {C}omputing in the {NISQ} era and beyond.
\newblock {\em {Quantum}}, 2:79, August 2018.

\bibitem{Feynman1996}
R.~P. Feynman.
\newblock {\em {Feynman lectures on gravitation}}.
\newblock Reading, Mass.: Addison-Wesley, 12 1996.

\bibitem{Diosi1987}
L.~Di{\'{o}}si.
\newblock A universal master equation for the gravitational violation of
  quantum mechanics.
\newblock {\em Phys. Lett. A}, 120(8):377--381, mar 1987.

\bibitem{Penrose1996}
R.~Penrose.
\newblock On gravity's role in quantum state reduction.
\newblock {\em Gen. Relativ. Gravit.}, 28(5):581--600, may 1996.

\bibitem{Bose2017}
S.~Bose, A.~Mazumdar, G.~W. Morley, H.~Ulbricht, M.~Toro\ifmmode~\check{s}\else
  \v{s}\fi{}, M.~Paternostro, A.~A. Geraci, P.~F. Barker, M.~S. Kim, and
  G.~Milburn.
\newblock Spin entanglement witness for quantum gravity.
\newblock {\em Phys. Rev. Lett.}, 119:240401, Dec 2017.

\bibitem{Blencowe2013}
M.~P. Blencowe.
\newblock Effective field theory approach to gravitationally induced
  decoherence.
\newblock {\em Phys. Rev. Lett.}, 111:021302, Jul 2013.

\bibitem{Pikovski2015}
I.~Pikovski, M.~Zych, F.~Costa, and {\v{C}}.~Brukner.
\newblock Universal decoherence due to gravitational time~dilation.
\newblock {\em Nat. Phys.}, 11(8):668--672, jun 2015.

\bibitem{Belenchia2019}
A.~Belenchia, R.~M. Wald, F.~Giacomini, E.~Castro-Ruiz, {\v{C}}.~Brukner, and
  M.~Aspelmeyer.
\newblock Information content of the gravitational field of a quantum
  superposition.
\newblock {\em Int. J. Mod. Phys. D}, 28(14):1943001, oct 2019.

\bibitem{Carney2019}
D.~Carney, P.~C.~E. Stamp, and J.~M. Taylor.
\newblock Tabletop experiments for quantum gravity: a user's manual.
\newblock {\em Class. Quantum Gravity}, 36(3):034001, jan 2019.

\bibitem{Asplemeyer2014}
M.~Aspelmeyer, T.~J. Kippenberg, and F.~Marquardt.
\newblock Cavity optomechanics.
\newblock {\em Rev. Mod. Phys.}, 86:1391--1452, Dec 2014.

\bibitem{Mancini1997}
S.~Mancini, V.~I. Man'ko, and P.~Tombesi.
\newblock Ponderomotive control of quantum macroscopic coherence.
\newblock {\em Phys. Rev. A}, 55:3042--3050, Apr 1997.

\bibitem{Giovannetti2001}
V.~Giovannetti, S.~Mancini, and P.~Tombesi.
\newblock Radiation pressure induced einstein-podolsky-rosen paradox.
\newblock {\em {EPL}}, 54(5):559--565, jun 2001.

\bibitem{Bose1997}
S.~Bose, K.~Jacobs, and P.~L. Knight.
\newblock Preparation of nonclassical states in cavities with a moving mirror.
\newblock {\em Phys. Rev. A}, 56:4175--4186, Nov 1997.

\bibitem{Mancini2002}
S.~Mancini, V.~Giovannetti, D.~Vitali, and P.~Tombesi.
\newblock Entangling macroscopic oscillators exploiting radiation pressure.
\newblock {\em Phys. Rev. Lett.}, 88:120401, Mar 2002.

\bibitem{Zhang2003}
J.~Zhang, K.~Peng, and S.~L. Braunstein.
\newblock Quantum-state transfer from light to macroscopic oscillators.
\newblock {\em Phys. Rev. A}, 68:013808, Jul 2003.

\bibitem{Pinrandola2006}
S.~Pirandola, D.~Vitali, P.~Tombesi, and S.~Lloyd.
\newblock Macroscopic entanglement by entanglement swapping.
\newblock {\em Phys. Rev. Lett.}, 97:150403, Oct 2006.

\bibitem{Corbitt2006}
T.~Corbitt, Y.~Chen, F.~Khalili, D.~Ottaway, S.~Vyatchanin, S.~Whitcomb, and
  N.~Mavalvala.
\newblock Squeezed-state source using radiation-pressure-induced rigidity.
\newblock {\em Phys. Rev. A}, 73:023801, Feb 2006.

\bibitem{Wipf2008}
C.~Wipf, T.~Corbitt, Y.~Chen, and N.~Mavalvala.
\newblock Route to ponderomotive entanglement of light via optically trapped
  mirrors.
\newblock {\em New J. Phys.}, 10(9):095017, sep 2008.

\bibitem{XinYou2015}
Xin-You L\"u, Jie-Qiao Liao, Lin Tian, and Franco Nori.
\newblock Steady-state mechanical squeezing in an optomechanical system via
  duffing nonlinearity.
\newblock {\em Phys. Rev. A}, 91:013834, Jan 2015.

\bibitem{Brooks2012}
D.~W.~C. Brooks, T.~Botter, S.~Schreppler, T.~P. Purdy, N.~Brahms, and D.~M.
  Stamper-Kurn.
\newblock Non-classical light generated by quantum-noise-driven cavity
  optomechanics.
\newblock {\em Nature}, 488(7412):476--480, aug 2012.

\bibitem{Purdy2013}
T.~P. Purdy, P.-L. Yu, R.~W. Peterson, N.~S. Kampel, and C.~A. Regal.
\newblock Strong optomechanical squeezing of light.
\newblock {\em Phys. Rev. X}, 3:031012, Sep 2013.

\bibitem{Aggarwal2018}
N.~Aggarwal, T.~Cullen, J.~Cripe, G.~D. Cole, R.~Lanza, A.~Libson, D.~Follman,
  P.~Heu, T.~Corbitt, and N.~Mavalvala.
\newblock Room temperature optomechanical squeezing.
\newblock arXiv: 1812.09942v1, 2018.

\bibitem{Chen2020}
J.~Chen, M.~Rossi, D.~Mason, and A.~Schliesser.
\newblock Entanglement of propagating optical modes via a mechanical interface.
\newblock {\em Nat. Commun.}, 11(1):943, feb 2020.

\bibitem{Murch2008}
K.~W. Murch, K.~L. Moore, S.~Gupta, and D.~M. Stamper-Kurn.
\newblock Observation of quantum-measurement backaction with an ultracold
  atomic gas.
\newblock {\em Nat. Phys.}, 4(7):561--564, may 2008.

\bibitem{Jayich2008}
A.~M. Jayich, J.~C. Sankey, B.~M. Zwickl, C.~Yang, J.~D. Thompson, S.~M.
  Girvin, A.~A. Clerk, F.~Marquardt, and J.~G.~E. Harris.
\newblock Dispersive optomechanics: a membrane inside a cavity.
\newblock {\em New J. Phys.}, 10(9):095008, sep 2008.

\bibitem{Painter2013}
A.~H. Safavi-Naeini, S.~Gröblacher, J.~T. Hill, J.~Chan, M.~Aspelmeyer, and
  O.~Painter.
\newblock Squeezed light from a silicon micromechanical resonator.
\newblock {\em Nature}, 500(7461):185--189, aug 2013.

\bibitem{Tian2013}
Lin Tian.
\newblock Robust photon entanglement via quantum interference in optomechanical
  interfaces.
\newblock {\em Phys. Rev. Lett.}, 110:233602, Jun 2013.

\bibitem{Rakhubovsky2020}
Andrey~A. Rakhubovsky, Darren~W. Moore, Uroš Delić, Nikolai Kiesel, Markus
  Aspelmeyer, and Radim Filip.
\newblock Detecting nonclassical correlations in levitated cavity
  optomechanics, 2020.

\bibitem{Barzanjeh2019}
S.~Barzanjeh, E.~S. Redchenko, M.~Peruzzo, M.~Wulf, D.~P. Lewis, G.~Arnold, and
  J.~M. Fink.
\newblock Stationary entangled radiation from micromechanical motion.
\newblock {\em Nature}, 570(7762):480--483, June 2019.

\bibitem{Wang2013}
Ying-Dan Wang and Aashish~A. Clerk.
\newblock Reservoir-engineered entanglement in optomechanical systems.
\newblock {\em Phys. Rev. Lett.}, 110:253601, Jun 2013.

\bibitem{Gut2020}
C.~Gut, K.~Winkler, J.~Hoelscher-Obermaier, S.~G. Hofer, R.~Moghadas Nia,
  N.~Walk, A.~Steffens, J.~Eisert, W.~Wieczorek, J.~A. Slater, M.~Aspelmeyer,
  and K.~Hammerer.
\newblock Stationary optomechanical entanglement between a mechanical
  oscillator and its measurement apparatus.
\newblock {\em Phys. Rev. Research}, 2:033244, Aug 2020.

\bibitem{Ockeloen-Korppi2018}
C.~F. Ockeloen-Korppi, E.~Damsk{\"a}gg, J.-M. Pirkkalainen, M.~Asjad, A.~A.
  Clerk, F.~Massel, M.~J. Woolley, and M.~A. Sillanp{\"a}{\"a}.
\newblock Stabilized entanglement of massive mechanical oscillators.
\newblock {\em Nature}, 556(7702):478--482, Apr 2018.

\bibitem{Palomaki2013}
T.~A. Palomaki, J.~D. Teufel, R.~W. Simmonds, and K.~W. Lehnert.
\newblock Entangling mechanical motion with microwave fields.
\newblock {\em Science}, 342(6159):710--713, October 2013.

\bibitem{Riedinger2016}
Ralf Riedinger, Sungkun Hong, Richard~A. Norte, Joshua~A. Slater, Juying Shang,
  Alexander~G. Krause, Vikas Anant, Markus Aspelmeyer, and Simon
  Gr\"{o}blacher.
\newblock Non-classical correlations between single photons and phonons from a
  mechanical oscillator.
\newblock {\em Nature}, 530(7590):313--316, January 2016.

\bibitem{Riedinger2018}
Ralf Riedinger, Andreas Wallucks, Igor Marinkovi{\'{c}}, Clemens
  L{\"o}schnauer, Markus Aspelmeyer, Sungkun Hong, and Simon Gr{\"o}blacher.
\newblock Remote quantum entanglement between two micromechanical oscillators.
\newblock {\em Nature}, 556(7702):473--477, Apr 2018.

\bibitem{Krisnanda2017}
T.~Krisnanda, M.~Zuppardo, M.~Paternostro, and T.~Paterek.
\newblock Revealing nonclassicality of inaccessible objects.
\newblock {\em Phys. Rev. Lett.}, 119:120402, Sep 2017.

\bibitem{Ma2020}
Yue Ma, Federico Armata, Kiran~E. Khosla, and M.~S. Kim.
\newblock Optical squeezing for an optomechanical system without quantizing the
  mechanical motion.
\newblock {\em Phys. Rev. Research}, 2:023208, May 2020.

\bibitem{Marletto2017}
C.~Marletto and V.~Vedral.
\newblock Gravitationally induced entanglement between two massive particles is
  sufficient evidence of quantum effects in gravity.
\newblock {\em Phys. Rev. Lett.}, 119:240402, Dec 2017.

\bibitem{DuanCriteria}
Lu-Ming Duan, G.~Giedke, J.~I. Cirac, and P.~Zoller.
\newblock Inseparability criterion for continuous variable systems.
\newblock {\em Phys. Rev. Lett.}, 84:2722--2725, Mar 2000.

\bibitem{Chang2009}
D.~E. Chang, C.~A. Regal, S.~B. Papp, D.~J. Wilson, J.~Ye, O.~Painter, H.~J.
  Kimble, and P.~Zoller.
\newblock Cavity opto-mechanics using an optically levitated nanosphere.
\newblock {\em Proc. Natl. Acad. Sci. U.S.A.}, 107(3):1005--1010, dec 2009.

\bibitem{Romero_Isart_2010}
O.~Romero-Isart, M.~L. Juan, R.~Quidant, and J.~I. Cirac.
\newblock Toward quantum superposition of living organisms.
\newblock {\em New J. Phys.}, 12(3):033015, mar 2010.

\bibitem{Kiesel2013}
N.~Kiesel, F.~Blaser, U.~Deli{\'c}, D.~Grass, R.~Kaltenbaek, and M.~Aspelmeyer.
\newblock Cavity cooling of an optically levitated submicron particle.
\newblock {\em Proc. Natl. Acad. Sci. U.S.A.}, 110(35):14180--14185, 2013.

\bibitem{Neumeier2018}
Lukas Neumeier, Tracy~E. Northup, and Darrick~E. Chang.
\newblock Reaching the optomechanical strong-coupling regime with a single atom
  in a cavity.
\newblock {\em Phys. Rev. A}, 97:063857, Jun 2018.

\bibitem{Brennecke2008}
F.~Brennecke, S.~Ritter, T.~Donner, and T.~Esslinger.
\newblock Cavity optomechanics with a bose-einstein condensate.
\newblock {\em Science}, 322(5899):235--238, oct 2008.

\bibitem{Takatsuji1967}
M.~Takatsuji.
\newblock Quantum theory of the optical kerr effect.
\newblock {\em Phys. Rev.}, 155:980--986, Mar 1967.

\bibitem{Lombardi2002}
Egilberto Lombardi, Fabio Sciarrino, Sandu Popescu, and Francesco De~Martini.
\newblock Teleportation of a vacuum--one-photon qubit.
\newblock {\em Phys. Rev. Lett.}, 88:070402, Jan 2002.

\bibitem{Guerreiro2016}
T.~Guerreiro, F.~Monteiro, A.~Martin, J.~B. Brask, T.~V\'ertesi, B.~Korzh,
  M.~Caloz, F.~Bussi\`eres, V.~B. Verma, A.~E. Lita, R.~P. Mirin, S.~W. Nam,
  F.~Marsilli, M.~D. Shaw, N.~Gisin, N.~Brunner, H.~Zbinden, and R.~T. Thew.
\newblock Demonstration of einstein-podolsky-rosen steering using single-photon
  path entanglement and displacement-based detection.
\newblock {\em Phys. Rev. Lett.}, 117:070404, Aug 2016.

\bibitem{Peyronel2012}
T.~Peyronel, O.~Firstenberg, Qi-Yu Liang, S.~Hofferberth, A.~V. Gorshkov,
  T.~Pohl, M.~D. Lukin, and V.~Vuleti{\'{c}}.
\newblock Quantum nonlinear optics with single photons enabled by strongly
  interacting atoms.
\newblock {\em Nature}, 488(7409):57--60, jul 2012.

\bibitem{Hofheinz2009}
M.~Hofheinz, H.~Wang, M.~Ansmann, R.~C. Bialczak, E.~Lucero, M.~Neeley, A.~D.
  O'Connell, D.~Sank, J.~Wenner, J.~M. Martinis, and A.~N. Cleland.
\newblock Synthesizing arbitrary quantum states in a superconducting resonator.
\newblock {\em Nature}, 459(7246):546--549, may 2009.

\bibitem{Hornberger2020}
Lukas Martinetz, Klaus Hornberger, James Millen, M.~S. Kim, and Benjamin~A.
  Stickler.
\newblock Quantum electromechanics with levitated nanoparticles, 2020.

\bibitem{Krantz2019}
P.~Krantz, M.~Kjaergaard, F.~Yan, T.~P. Orlando, S.~Gustavsson, and W.~D.
  Oliver.
\newblock A quantum engineer's guide to superconducting qubits.
\newblock {\em Appl. Phys. Rev.}, 6(2):021318, June 2019.

\bibitem{HeinzPeterBreuer2007}
Heinz-Peter Breuer and F.~Petruccione.
\newblock {\em The Theory of Open Quantum Systems}.
\newblock Oxford University Press, 2007.

\bibitem{Gaussian_Quantum_Information}
C.~Weedbrook, S.~Pirandola, R.~Garc\'{\i}a-Patr\'on, N.~J. Cerf, T.~C. Ralph,
  J.~H. Shapiro, and S.~Lloyd.
\newblock Gaussian quantum information.
\newblock {\em Rev. Mod. Phys.}, 84:621--669, May 2012.

\bibitem{Davidovich}
Luiz Davidovich.
\newblock Sub-poissonian processes in quantum optics.
\newblock {\em Rev. Mod. Phys.}, 68:127--173, Jan 1996.

\bibitem{Lloyd1999}
Seth Lloyd and Samuel~L. Braunstein.
\newblock Quantum computation over continuous variables.
\newblock {\em Phys. Rev. Lett.}, 82:1784--1787, Feb 1999.

\bibitem{Kippenberg2004}
T.~J. Kippenberg, S.~M. Spillane, and K.~J. Vahala.
\newblock Kerr-nonlinearity optical parametric oscillation in an ultrahigh-$q$
  toroid microcavity.
\newblock {\em Phys. Rev. Lett.}, 93:083904, Aug 2004.

\bibitem{Rabl2011}
P.~Rabl.
\newblock Photon blockade effect in optomechanical systems.
\newblock {\em Phys. Rev. Lett.}, 107:063601, Aug 2011.

\bibitem{Meyer2019}
Nadine Meyer, Andr\'es de los~Rios Sommer, Pau Mestres, Jan Gieseler, Vijay
  Jain, Lukas Novotny, and Romain Quidant.
\newblock Resolved-sideband cooling of a levitated nanoparticle in the presence
  of laser phase noise.
\newblock {\em Phys. Rev. Lett.}, 123:153601, Oct 2019.

\bibitem{Delic2019}
U.~Deli{\'c}, Manuel Reisenbauer, David Grass, Nikolai Kiesel, Vladan
  Vuleti\ifmmode~\acute{c}\else \'{c}\fi{}, and Markus Aspelmeyer.
\newblock Cavity cooling of a levitated nanosphere by coherent scattering.
\newblock {\em Phys. Rev. Lett.}, 122:123602, Mar 2019.

\bibitem{Delic892}
Uro{\v s} Deli{\'c}, Manuel Reisenbauer, Kahan Dare, David Grass, Vladan
  Vuleti{\'c}, Nikolai Kiesel, and Markus Aspelmeyer.
\newblock Cooling of a levitated nanoparticle to the motional quantum ground
  state.
\newblock {\em Science}, 367(6480):892--895, 2020.

\bibitem{Windey2019}
Dominik Windey, Carlos Gonzalez-Ballestero, Patrick Maurer, Lukas Novotny,
  Oriol Romero-Isart, and Ren\'e Reimann.
\newblock Cavity-based 3d cooling of a levitated nanoparticle via coherent
  scattering.
\newblock {\em Phys. Rev. Lett.}, 122:123601, Mar 2019.

\bibitem{Paternostro2007}
M.~Paternostro, D.~Vitali, S.~Gigan, M.~S. Kim, C.~Brukner, J.~Eisert, and
  M.~Aspelmeyer.
\newblock Creating and probing multipartite macroscopic entanglement with
  light.
\newblock {\em Phys. Rev. Lett.}, 99:250401, Dec 2007.

\bibitem{RiosSommer2020}
A.~de~los Ríos~Sommer, N.~Meyer, and R.~Quidant.
\newblock Strong optomechanical coupling at room temperature by coherent
  scattering.
\newblock arXiv: 2005.10201v1, 2020.

\bibitem{Hornberger2020_CS}
Henning Rudolph, Klaus Hornberger, and Benjamin~A. Stickler.
\newblock Entangling levitated nanoparticles by coherent scattering.
\newblock {\em Phys. Rev. A}, 101:011804, Jan 2020.

\bibitem{Radim2020}
Ond\ifmmode \check{r}\else~\v{r}\fi{}ej \ifmmode~\check{C}\else
  \v{C}\fi{}ernot\'{\i}k and Radim Filip.
\newblock Strong mechanical squeezing for a levitated particle by coherent
  scattering.
\newblock {\em Phys. Rev. Research}, 2:013052, Jan 2020.

\bibitem{Chauhan2020}
Anil~Kumar Chauhan, Ondřej Černotík, and Radim Filip.
\newblock Stationary gaussian entanglement between levitated nanoparticles,
  2020.

\bibitem{Gieseler2012}
Jan Gieseler, Bradley Deutsch, Romain Quidant, and Lukas Novotny.
\newblock Subkelvin parametric feedback cooling of a laser-trapped
  nanoparticle.
\newblock {\em Phys. Rev. Lett.}, 109(10), sep 2012.

\bibitem{Conangla2019}
Gerard~P. Conangla, Francesco Ricci, Marc~T. Cuairan, Andreas~W. Schell, Nadine
  Meyer, and Romain Quidant.
\newblock Optimal feedback cooling of a charged levitated nanoparticle with
  adaptive control.
\newblock {\em Phys. Rev. Lett.}, 122(22), jun 2019.

\end{thebibliography}

\end{document}